\documentclass[%
 aip,
 amsmath,amssymb,
 reprint,%
]{revtex4-1}

\usepackage{graphicx}
\usepackage{dcolumn}
\usepackage{bm}

\usepackage{cleveref}

\usepackage[utf8]{inputenc}
\usepackage[T1]{fontenc}
\usepackage{mathptmx}
\usepackage{etoolbox}
\usepackage{natbib}
 
\usepackage{xcolor}

\usepackage[normalem]{ulem}
 
\makeatletter
\def\@email#1#2{%
 \endgroup
 \patchcmd{\titleblock@produce}
  {\frontmatter@RRAPformat}
  {\frontmatter@RRAPformat{\produce@RRAP{*#1\href{mailto:#2}{#2}}}\frontmatter@RRAPformat}
  {}{}
}%
\makeatother
\begin{document}

\preprint{AIP/123-QED}

\title{Synthesizing impurity clustering in the edge plasma of tokamaks using neural networks}

\author{Z. Lin}
\affiliation{ 
 Aix-Marseille Universit\'e, CNRS, I2M, UMR 7373, 13453 Marseille, France}%
 \author{T. Maurel--Oujia}
 %Orcid number: 0000-0002-2806-7157
\affiliation{ 
 Aix-Marseille Universit\'e, CNRS, I2M, UMR 7373, 13453 Marseille, France}%
\author{B. Kadoch}
%Orcid number: 0000-0001-9346-1399
\affiliation{ 
 Aix-Marseille Universit\'e, CNRS, IUSTI, UMR 7343, 13453 Marseille, France}%
 \author{P. Krah}
 %Orcid number: 0000-0002-2806-7157
\affiliation{ 
 Aix-Marseille Universit\'e, CNRS, I2M, UMR 7373, 13453 Marseille, France}%
 \author{N. Saura}
\affiliation{ 
Aix-Marseille Universit\'e, CNRS, PIIM, UMR 7345, 13397 Marseille, France}%
 \author{S. Benkadda}
\affiliation{ 
Aix-Marseille Universit\'e, CNRS, PIIM, UMR 7345, 13397 Marseille, France}

\author{K. Schneider}%
 \email{kai.schneider@univ-amu.fr}
\affiliation{ 
 Aix-Marseille Universit\'e, CNRS, I2M, UMR 7373, 13453 Marseille, France}%
 %Orcid number: 0000-0003-1243-6621

\date{\today}% It is always \today, today,

\begin{abstract}
 This work investigates the behavior of impurities in edge plasma of tokamaks using high-resolution numerical simulations based on Hasegawa--Wakatani equations. Specifically, it focuses on the behavior of inertial particles, which has not been extensively studied in the field of plasma physics. Our simulations utilize one-way coupling of a large number of inertial point particles, which model plasma impurities. We observe that with Stokes number ($St$) which characterizes the inertia of particles being much less than one, such light impurities closely track the fluid flow without pronounced clustering. For intermediate $St$ values, distinct clustering appears, with larger Stokes values, {\it i.e.} heavy impurities even generating more substantial clusters. When $St$ is significantly large, very heavy impurities tend to detach from the flow and maintain their trajectory, resulting in fewer observable clusters and corresponding to random motion. A core component of this work involves machine learning techniques. Applying three different neural networks - Autoencoder, U-Net, and  Generative Adversarial Network (GAN) - to synthesize preferential concentration fields of impurities, we use vorticity as input and predict impurity number density fields. GAN outperforms the two others by aligning closely with direct numerical simulation data in terms of probability density functions of the particle distribution and energy spectra. This machine learning technique holds the potential to reduce computational costs by eliminating the need to track millions of particles modeling impurities in simulations.\\ 
\end{abstract}

\maketitle

\section{Introduction}
The pursuit of a sustainable and abundant energy source has sparked greater attention towards nuclear fusion, which is the process responsible for powering celestial bodies such as stars. Nuclear fusion is distinct from nuclear fission as it involves the fusion of light atomic nuclei to create heavier ones, resulting in the release of a significant amount of energy. Fusion is highly regarded as the ultimate goal in energy production due to its capacity to generate significant power while causing minimal harm to the environment.

Fusion reactors confine and heat a plasma, consisting of ions and electrons, to enable fusion reactions at high temperatures. However, the task of confining plasma poses a considerable challenge due to the presence of multiple instabilities that can result in energy loss. Impurity accumulation in the plasma core can result in heat loss through radiation, leading to a decrease in confinement quality. Therefore, it is crucial to conduct a thorough investigation of impurity in fusion plasma.

The dynamics of magnetically confined plasma flow are influenced by drift-wave turbulence and zonal flows. The numerical high-resolution simulations for modeling the plasma flow carried out within this work are based on the Hasegawa--Wakatani model, which governs cross-field transport by electrostatic drift waves in magnetically confined plasmas. A modified version of the model\cite{Pushkarev2013} is likewise investigated to take into account zonal flows. The electric field perpendicular to magnetic field lines is particularly significant because it strongly drives cross-field fluxes, impacting edge pressure profiles and stability.\cite{Wang2020,  Zhang2020} $\boldsymbol{E} \times \boldsymbol{B}$ flows strongly influence the motion of coherent structures, which can account for substantial particle losses in tokamaks.\cite{Carralero2017, DIppolito2011, Kube2018} The resulting fluxes on walls can cause sputtering, erosion, and impurity injection, further degrading confinement and safe operation.\cite{DIppolito2011, Kuang2018, Kuang2020}

For the particles, there has been growing interest in Lagrangian perspective in recent years.\cite{gheorghiu2024transport,Kadoch2022, Bos2010} This approach involves the analysis of trajectories of numerous tracers. Numerical simulations involve solving equations to determine the trajectories of test particles within a specific velocity field, such as the $\boldsymbol{E} \times \boldsymbol{B}$ velocity field. The Lagrangian approach highlights the significant impact of coherent structures on transport. The combined effect of eddy trapping and zonal shear flows often leads to non-diffusive transport. \cite{Krasheninnikov2008, delCastilloNegrete2004, vanMilligen2004} In fusion plasma research, previous studies have primarily focused on passive flow tracers without considering inertial effects in the edge plasma using the Hasegawa--Wakatani model.\cite{Futatani2008, Futatani2009a, Futatani2008a}  We increase the complexity of these models and add the effect of inertia, which has not been done so far, to the best of our knowledge. Therewith we can study the behavior of heavy and light atoms and see the impact on the impurity distribution. A crucial aspect of understanding inertial impurities is their self-organization, such as clustering and void formation, which can be quantified mathematically by deviation from Poissonian statistics.\cite{Oujia2020} 
Their dynamics can then be analyzed using tesselation based techniques.\cite{maurel2024computing}

There is a growing interest in studying spatial patterns in turbulence exploring machine learning.\cite{Ishikawa2022, Jajima2023} Our study uses machine learning to estimate mesoscale inertial particle clustering in turbulence, relying on flow field data rather than individual particle information. We apply machine learning tools that have been successfully implemented in the field of hydrodynamics, see Ref.~\onlinecite{Oujia2022} and \onlinecite{Maurel2023NN}. Three neural networks — Autoencoder, U-Net, and Generative Adversarial Network (GAN) — are trained to predict impurity particle densities from vorticity field snapshots. The results are visually and statistically compared with actual DNS particle data. This approach reduces the cost for predicting particle density distributions.

The paper is structured as follows: Sec.~\ref{sec: Models} outlines the theoretical basis used for simulation, introducing the Hasegawa--Wakatani edge plasma model and the motion of inertial impurity particles. Sec.~\ref{sec: Numerical Simulations} details the numerical simulations and results. In Sec.~\ref{sec: Neural Networks}, we explore machine learning models such as Autoencoders, U-net, and GAN to synthesize impurity concentration fields. Finally, Sec.~\ref{sec: Conclusions} summarizes our findings and suggests potential directions for future investigations.

\section{\label{sec: Models}Models for Edge Plasma Turbulence and impurity particles}\subsection{Hasegawa--Wakatani model}

\begin{figure}[htb]
        \centerline{\includegraphics[width=0.5\textwidth]{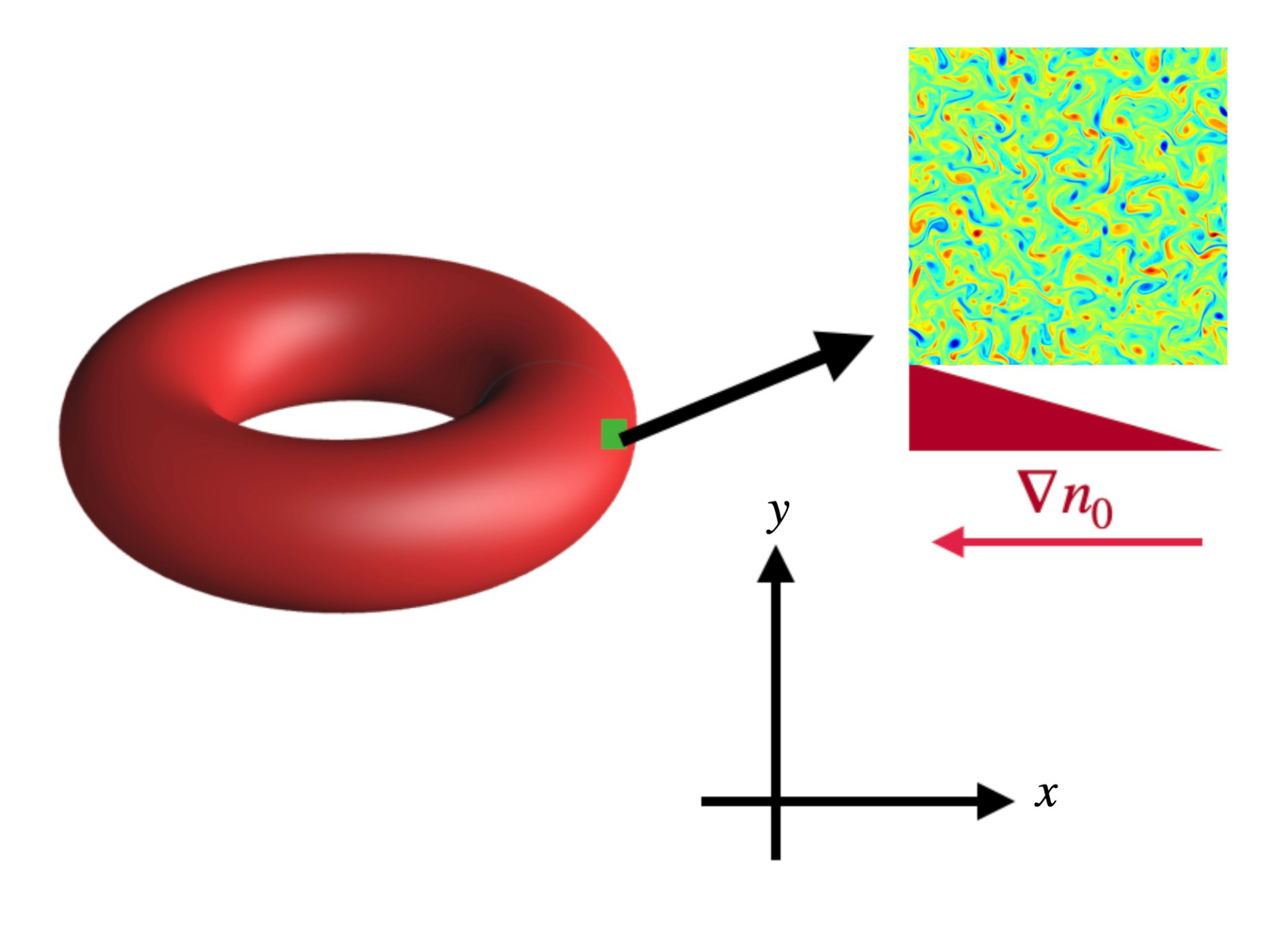}}
        \caption[]{Illustration of the two-dimensional slab geometry in the tokamak edge used in the Hasegawa--Wakatani system. The radial direction is represented by $x$, while $y$ is the poloidal direction. There is an imposed mean plasma density gradient $\nabla{n_0}$ in the radial direction. The 2D flow is computed within a domain whose size is equivalent to 64 times the Larmor radius ($\rho_s$). This computational domain is marked with a green square. On the right, a vorticity field for the classical Hasegawa–Wakatani (cHW) is shown. The figure is adapted from Ref.~\onlinecite{Bos2010}.}.
        \label{fig: slab geometry}
 \end{figure}

The numerical simulations are based on Hasegawa--Wakatani (HW) equations for plasma edge turbulence driven by drift-wave instability.\cite{Hasegawa1983} Our focus is on the two-dimensional, slab geometry of the HW model, see {\it e.g.} Ref.~\onlinecite{Kadoch2022}.  Fig.~\ref{fig: slab geometry} depicts a representation of the flow configuration. The magnetic field lines are assumed to be constant,  straight and perpendicular to the slab. Ions are considered cold, and the effects of the temperature gradient are ignored.

The HW model consists of two partial differential equations that describe the time evolution of the plasma electrostatic potential $\phi$ and fluctuating plasma density $n$:

\begin{equation}
\left(\frac{\partial}{\partial t}- \nu \nabla^{2}\right) \nabla^{2} \phi=\left[\nabla^{2} \phi, \phi\right]+c(\phi-n),
\label{eq: phi}
\end{equation}

\begin{equation}
\left(\frac{\partial}{\partial t}-D \nabla^{2}\right) n=[n, \phi]-\kappa \frac{\partial \phi}{\partial y}+c(\phi-n),
\label{eq: n}
\end{equation} 

All quantities are dimensionless and normalized as described in Refs.~\onlinecite{Bos2010, Futatani2011}. The constant $D$  and $\nu$ are cross-field diffusion coefficient of plasma density fluctuations and kinematic viscosity, respectively. The constant $\kappa$ defined as $\kappa \equiv - \partial_{x} \ln n_{0}$, is a measure of the density gradient. The Poisson bracket is defined as: $[A, B]=\frac{\partial A}{\partial x} \frac{\partial B}{\partial y}-\frac{\partial A}{\partial y} \frac{\partial B}{\partial x}$. In this system, the electrostatic potential $\phi$ acts as a stream-function for the $\boldsymbol{E}  \times \boldsymbol{B}$ flow velocity, denoted by $\boldsymbol{u}=\nabla^{\perp} \phi$, where $\nabla^{\perp}=\left(-\partial_y, \partial_x\right)$. Thus, we have $u_x=-\partial \phi / \partial y$ and $u_y=\partial \phi / \partial x$. The vorticity is defined as $\omega=\nabla^2 \phi$. The turnover time of turbulent eddies, $\tau_{\eta}$, is defined as $1 / \sqrt{2 Z_{ m s}}$, with $Z_{ m s}$ denoting the mean-square vorticity.  The adiabaticity parameter $c$ is:

\begin{equation}
c=\frac{T_{e} k_{\parallel}^{2}}{\mathrm{e}^{2} n_{0} \eta \omega_{c i}}
\end{equation}

where $\eta$ is the electron resistivity.  Here $k_{\parallel}$ is the effective parallel wavenumber. The parameter $c$ controls the phase difference between the electrostatic potential and the plasma density fluctuations.  The case $c\gg 1$ is known as the adiabatic limit. In this limit, the Hasegawa--Wakatani model reduces to the Hasegawa--Mima equation, and the electrons in the plasma follow a Boltzmann distribution.  In that case, there is no phase shift between $\phi$ and $n$. Conversely, when $c \ll 1$ (the hydrodynamic limit), the system reduces to a form that is analogous to a two-dimensional Navier--Stokes equation. In this regime, the density fluctuations are passively advected by the $\boldsymbol{E} \times \boldsymbol{B}$  flow.\cite{Kadoch2022}  The most  physically interesting situation in fusion plasma is when $c \sim 1$, known as the quasi-adiabatic regime. In this regime, there is a phase shift between $\phi$ and $n$. This phase shift permits the presence of particle transport.

In the context of the classical Hasegawa--Wakatani model (cHW), as detailed above, zonal flows are not observed. To specifically obtain zonal flows, a  modified Hasegawa--Wakatani model (mHW)  can be considered. This modification was introduced in Ref.~\onlinecite{Pushkarev2013} where the coupling term $c(\phi -n)$ is set to zero for modes $k_y= 0$.

\subsection{Impurity particles model}
Previous studies \cite{Futatani2008, Futatani2009a, Futatani2008a} have assumed that impurity particles perfectly follow the fluid flow without deviating from the flow path.  However, this assumption may not always be valid, particularly when the particles have significant mass, resulting in observable inertial effects.

 \subsubsection{Stokes number}
To account for the inertial effects, the concept of the Stokes number ($St$) is introduced. \cite{Maxey1987}  The Stokes number is a dimensionless parameter used to quantify the inertia of particles in a fluid flow. The Stokes number is defined as:

\begin{equation}
St = \frac{\tau_{p}}{\tau_{\eta}},
\end{equation}
where $\tau_{{p}}$ is the impurity particle's relaxation time. This is a measure of how quickly an impurity particle can respond to changes in the fluid flow. It is determined by factors like the particle's size, mass, as well as the viscosity of the fluid. The turbulent eddy turnover time is $\tau_{\eta}$. This is a measure of how quickly the fluid flow is changing.   A high Stokes number suggests that the particle's motion is dominated by its inertia, meaning it will continue moving in its current direction even when the fluid flow changes.  Note that the Stokes number is a simplification, and real-world particle-fluid interactions may involve other complex phenomena, like particle-particle interactions, two-way coupling, finite particle-size effects, etc. Nevertheless, the Stokes number provides a valuable prediction of particle behavior in a turbulent flow.

\subsubsection{Motion of inertial impurity particles}

We assume the behavior of impurities does not influence the plasma's overall dynamics. This study is focused on tracking individual particles using Lagrangian mechanics. The equations governing the motion of these particles are:
\begin{equation}
\frac{d \boldsymbol{x}_{imp,j}}{dt} = \boldsymbol{v}_{imp,j}  
\label{eq: x}
\end{equation}

\begin{equation}
\frac{d \boldsymbol{v}_{imp,j}}{dt} = -\frac{\boldsymbol{v}_{imp,j} - \boldsymbol{u}_{imp,j}}{\tau_{p}}
\label{eq: v}
\end{equation}
The particles are denoted by $j$ ($j=1, \ldots, N$, where $N$ is the total number of impurity particles in the system). Here, $\boldsymbol{v}_{imp,j}$ is the velocity vector of the $j$-th impurity particle, and $\boldsymbol{u}_{imp,j}$ represents the fluid velocity vector at the location of the particle, $\boldsymbol{x}_{imp,j}$. The force resisting the motion of these particles is assumed to be in the form of Stokes drag, which is directly proportional to the relative velocity between the particle and the fluid. The constant of proportionality, $1/\tau_{p}$, is the inverse of particle relaxation time.\cite{Onishi2011, Obligado2014, Matsuda2019} In the initial phase of our study, we excluded electromagnetic forces to focus on establishing a fundamental understanding without introducing additional complexity. The effects of electromagnetic forces on the particles will be integrated incrementally in future studies.

\section{\label{sec: Numerical Simulations}Numerical Simulations} 

\subsection{Flow configurations and physical parameters}

The simulation was conducted within a domain with periodic boundary condition that spans an area of $64^2$ and it was discretized using a resolution of $1024^2$ grid points. The time step was $5\times10^{-4}$, while values for diffusivity of plasma density $D$ and kinematic viscosity $\nu$ were both set to $5\times10^{-3}$.  The value for $\kappa$ was set to 1. Simulations for the flow begin with Gaussian random initial conditions and run until achieving a saturated, fully developed turbulent flow (see Ref.~\onlinecite{Kadoch2022}). After this phase,$10^6$ uniformly distributed inertial impurity particles are injected.

\subsection{Numerical schemes}
The systems of HW equations, specifically Eq.~\eqref{eq: phi} and Eq.~\eqref{eq: n} are solved using a Fourier pseudospectral method. It is particularly well-suited for problems with periodic boundary conditions and provides excellent accuracy for smooth solutions.\cite{canuto2007spectral} The equations governing the motion of the impurity particles, namely Eq.~\eqref{eq: x} and Eq.~\eqref{eq: v}, are solved using a second-order Runge--Kutta (RK2) scheme. This method provides a balance between accuracy and computational efficiency. Simulating a large ensemble of particles presents a considerable computational challenge. Our solution leverages High-Performance Computing (HPC) to overcome this complexity, employing parallel computing with Message Passing Interface (MPI). For further details on the numerical method and its implementation, we refer to Ref.~\onlinecite{Kadoch2022}.

\subsection{Results}

\begin{figure*}[htb]
{\includegraphics[width=1.0\linewidth]{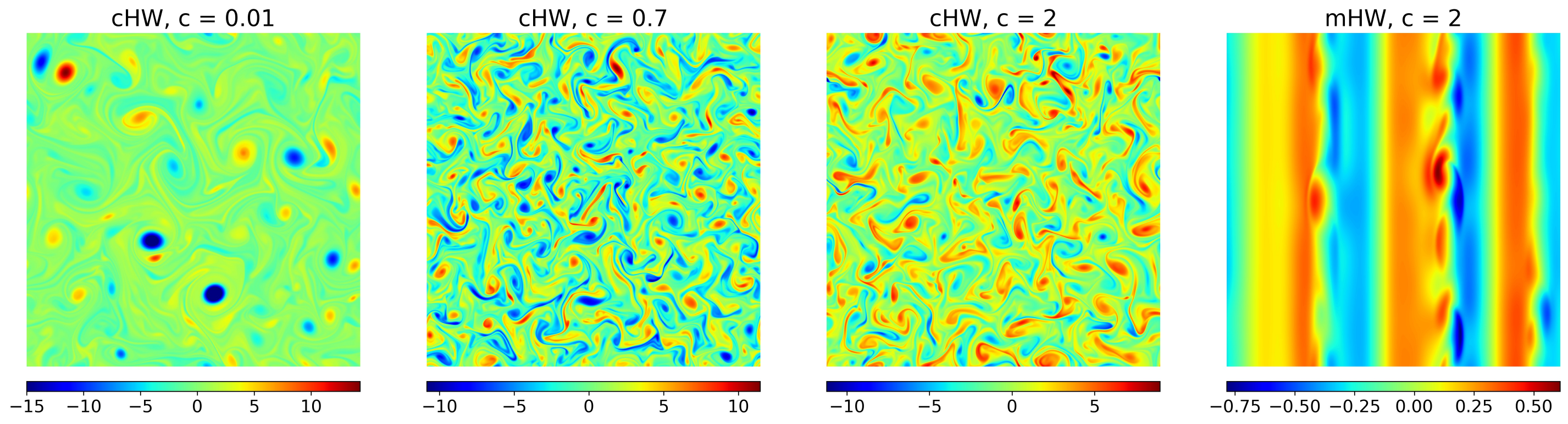}}
    \caption{Vorticity fields without impurity particles in fully developed turbulence regime. }
    \label{fig: mHW, vorticity field}
\end{figure*}

\begin{figure}[htb]
        \centerline{\includegraphics[width=0.5\textwidth]{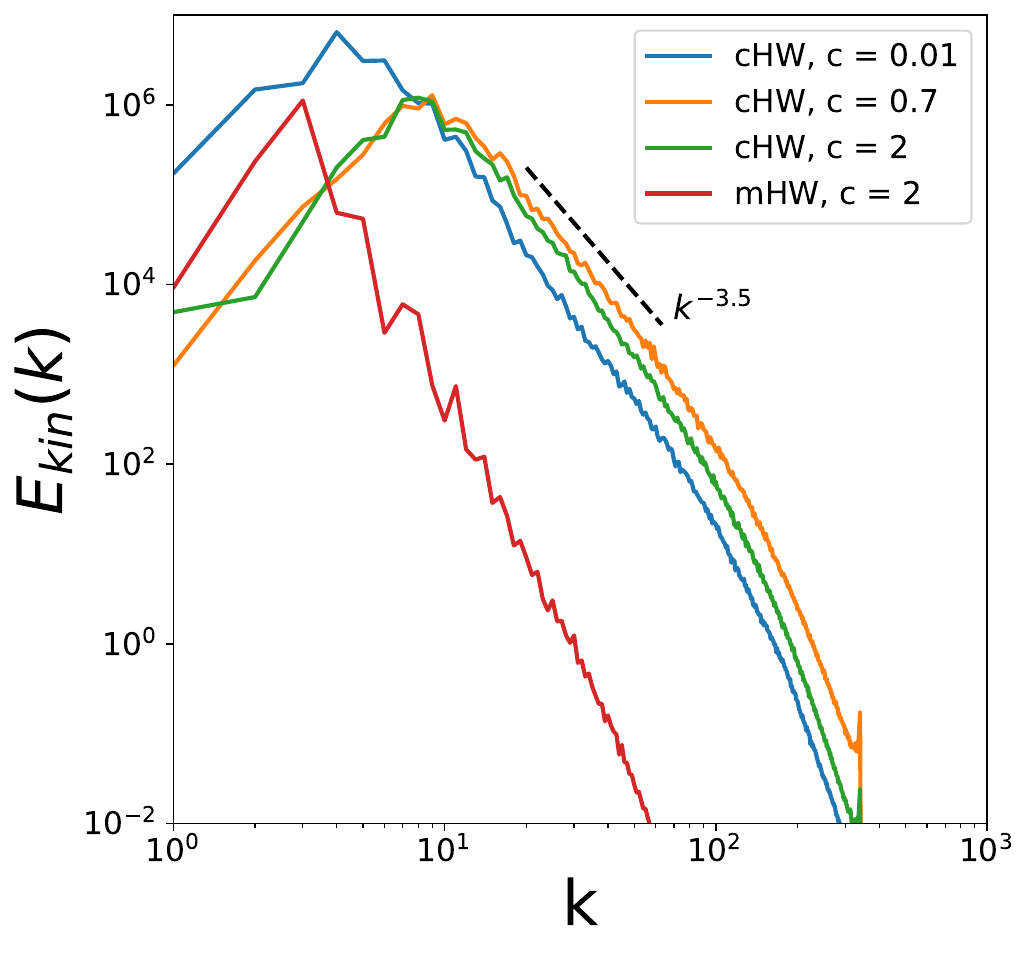}}
        \caption[]{ Kinetic energy spectra of the flow. The slope $k^{-3.5}$ is provided for reference.   }

        \label{fig: energy_spectra}
 \end{figure}
Fig.~\ref{fig: mHW, vorticity field} shows the vorticity field (without impurity particles) in a fully developed turbulence regime of classical Hasegawa--Wakatani model (cHW) for different $c$ values and modified Hasegawa--Wakatani model (mHW) for $c =2$. Larger vortices were exhibited in hydrodynamic cases (cHW, $c = 0.01$), while smaller ones were observed in quasi-adiabatic (cHW, $c = 0.7$) and adiabatic cases (cHW, $c = 2$). For the modified Hasegawa--Wakatani model, with $c = 2$, zonal flows are presented. To  quantify the distribution of kinetic energy of the flow across different scales of the turbulent flow, we compute the Fourier energy spectra:\cite{bassenne2017}

\begin{equation}
    E(k) = \frac{1}{2} \, \sum_{k - \frac{1}{2} \le|{\boldsymbol k}| < k + \frac{1}{2}} |\widehat {\boldsymbol u}({\boldsymbol k})|^2
    \label{eq:enspec_u}
\end{equation}
 
where $\widehat {\boldsymbol u}({\boldsymbol k})  = \int \int\, {\boldsymbol u}({\boldsymbol x}) \exp(- i 2 \pi {\boldsymbol k} \cdot {\boldsymbol x} ) \, d{\boldsymbol x}$ is the 2D Fourier transform of the flow velocity, $i = \sqrt{-1}$ and ${\boldsymbol k} = (k_x, k_y)$ the wave vector. The summation in Eq.~(\ref{eq:enspec_u}) is carried out over concentric shells in wave number space. In Fig.~\ref{fig: energy_spectra} the  Fourier energy spectra $E(k)$ are shown. We observe a peak at smaller wavenumbers for the cases $c = 0.01$ (cHW and mHW) and $c = 2$ (mHW), indicating the presence of larger structures. Furthermore, the spectra display a power-law scaling, with an exponent nearing -3.5 for the case $c = 0.7$.

In this paper, we focus on the quasi-adiabatic regime ($c = 0.7$), which is the relevant case for edge plasma of tokamaks. For readers interested in zonal flow, relevant discussions can be found in Appendix~\ref{appendix: zonal flows}. Fig.~\ref{fig: vorticity and particles: cHW, c =0.7} shows the visualization of vorticity fields and $10^4$ superimposed impurity particles for various Stokes numbers in quasi-adiabatic regime. From Fig.~\ref{fig: vorticity and particles: cHW, c =0.7} we can observe clustering of impurity particles for the case of $St =  0.25$ and  $St = 1$, while for $St =0$ and $St = 50$, the impurity density field homogeneously fills the entire physical domain and does not show significant spatial correlation. When the particles are very light, the response time of the particles is much smaller than the turnover time of the turbulent eddies, i.e., $\tau_p \ll \tau_{\eta}$ ($St \sim  0$). Their response time is so quick that they can adjust their velocity to match any changes in the flow immediately. They behave as usual tracers like those in previous studies. \cite{Futatani2008, Futatani2009a, Futatani2008a} When $\tau_p \gg \tau_{\eta}$ ($St \gg 1$), the inertia of the impurity particles is so high that they are almost unaffected by the eddies. Particles are moving ballistically and are randomly distributed, resulting in less clustering and more dispersion. When the response time of the particles is comparable to the eddy turnover time, i.e., $\tau_p \sim \tau_{\eta}$ ($St \sim 1$), particles are subject to centrifugal effect of turbulent eddies. Eddies can accelerate the particles to a degree where they are thrown out of the eddy due to their inertia. This ejection and the subsequent movement towards areas of lower vorticity result in particle clustering.  The clustering of impurity particles results in inhomogeneous
poloidal distributions ($y$ direction in the domain) of the impurity density, which can modify radial collisional transport.\cite{angioni2021}

 \begin{figure*}[htb]
        \centerline{\includegraphics[width=1.0\linewidth]{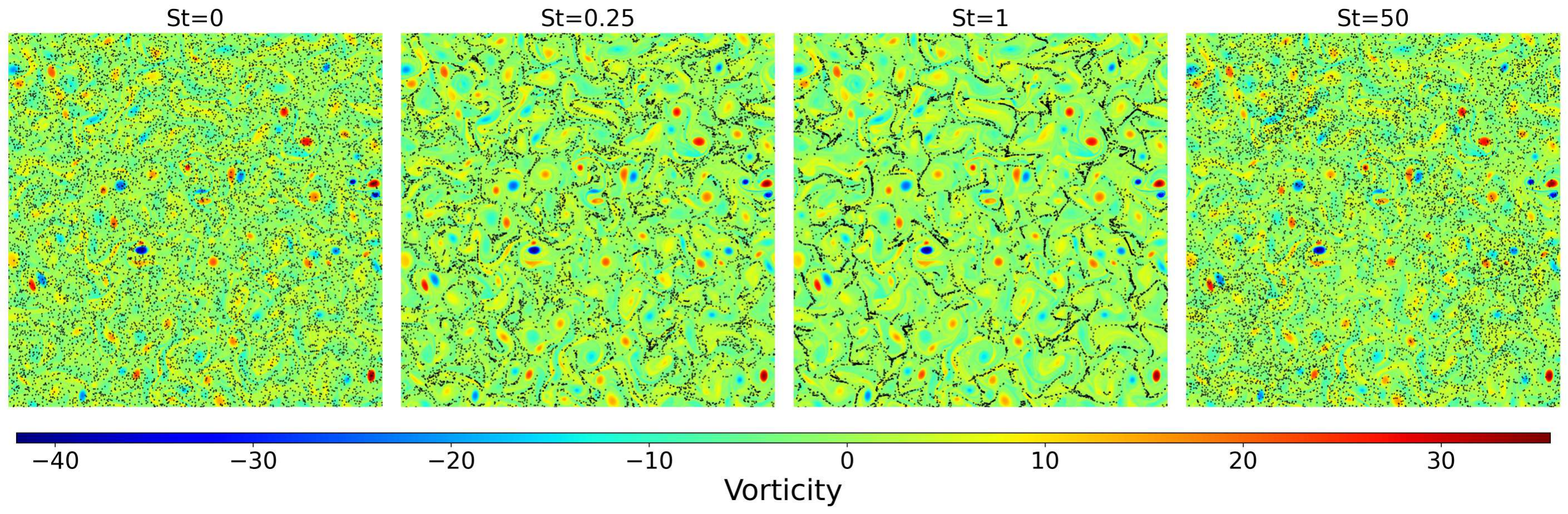}}
        \caption[]{Visualization of vorticity fields in fully developed turbulence regime and superimposed impurity particles ($10^4$ impurity particles) for various Stokes numbers in the case of $c = 0.7$ (cHW).  
     }
        \label{fig: vorticity and particles: cHW, c =0.7}
 \end{figure*}

To analyze and statistically interpret the distribution of impurity particles, we increase the number of particles from $10^4$ to $10^6$.   Displaying the position of each particle in the flow is impractical due to this large quantity. Instead, we visualize the particle density. We achieve this by dividing the entire domain into $512 \times 512$ boxes and defining the impurity density $n_{imp}$ as the number of particles in each box, i.e. we compute a histogram.  Fig.~\ref{fig: impurity density fields: cHW, c =0.7} shows the impurity density field for the case of  $c = 0.7$ considering different Stokes numbers. When $St = 0$, the impurity density is distributed uniformly throughout the flow.  As $St$ increases, variations in impurity density become apparent, with some areas exhibiting high density and others low. This uneven distribution indicates the formation of voids and clustering of particles. When $St=50$, the impurity density returns to a more uniform distribution. The impurity density fields exhibit patterns similar to those seen in the earlier visualization involving $10^4$ impurity particles.

\begin{figure}[htb]
      \centerline{\includegraphics[width=0.5\textwidth]{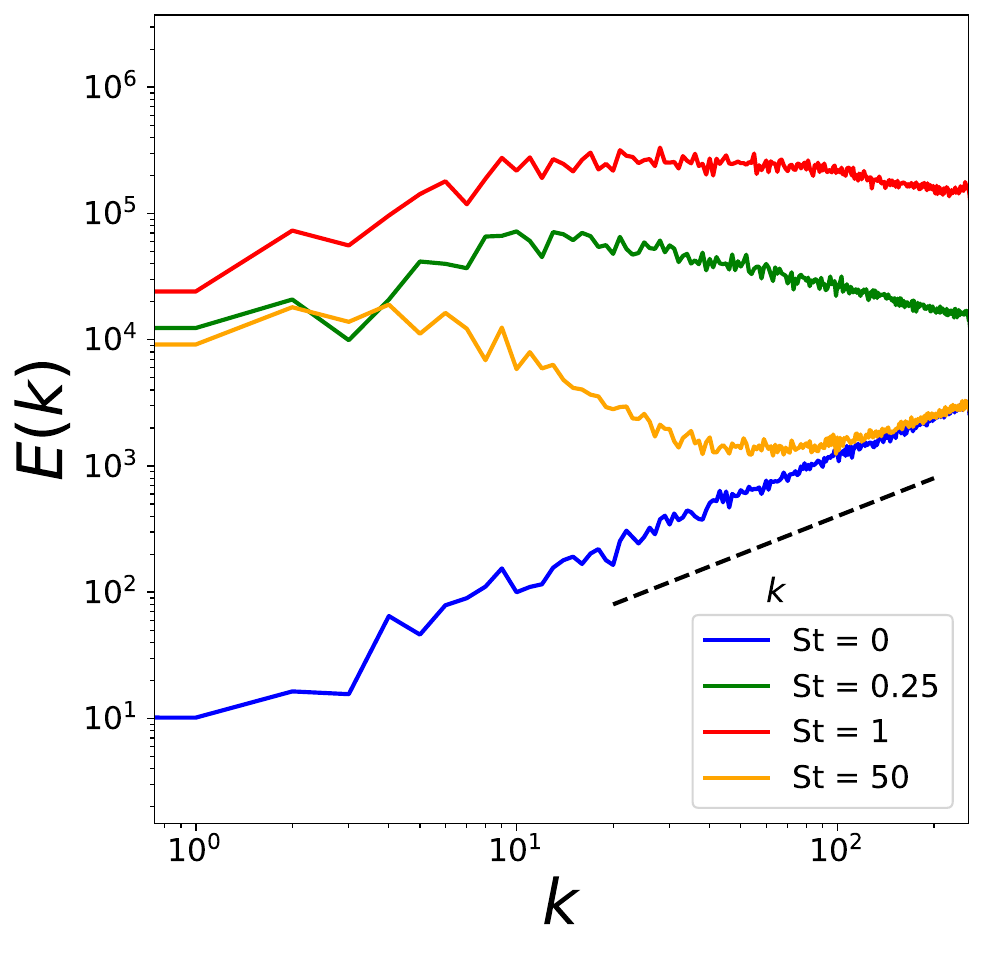}}
        \caption[]{Energy spectra of impurity density for $St$ = 0, 0.25, 1 and 50 in the case of $c = 0.7$(cHW). The slope $k$ is provided for reference. }
        \label{fig: density energy spectra}
 \end{figure}

To analyze the impurity density distribution at different scales, we compute the Fourier energy spectra  similar to Eq.~(\ref{eq:enspec_u}) replacing  the flow velocity ${\boldsymbol u}$ by the impurity density distribution $n_{imp}$. The spectral characteristics of impurity density field are shown in Fig.~\ref{fig: density energy spectra} in terms of the Fourier energy spectra. The variance of the impurity density is equivalent to the cumulative sum of the corresponding spectral energy. This is visually represented as the area under the energy spectra curve. As observed, the variance initially increases with the Stokes number up to $St = 1$, indicating stronger fluctuations. From $St = 1$ to $St = 50$, there is a decrease, suggesting less fluctuations. For $St=0$, representing tracer particles without inertia as in previous studies, the impurity particles are distributed randomly. Random spatial distribution of particles follows a Poisson probability density function (PDF) with mean and variance equal to the mean density $\langle n_{imp }\rangle$, see e.g. Ref.~\onlinecite{bassenne2017}. The slope of the spectra for $St = 0$ in 2D is 1, as prescribed by the equidistribution of variance among all wavenumbers. The spectrum for $St = 50 $ is observed to collapse on those of  $St = 0$ at small scales (high wavenumbers), thus resembling the same characteristics as for randomly distributed particles.

 \begin{figure*}[htb]
        \centerline{\includegraphics[width=1.0\linewidth]{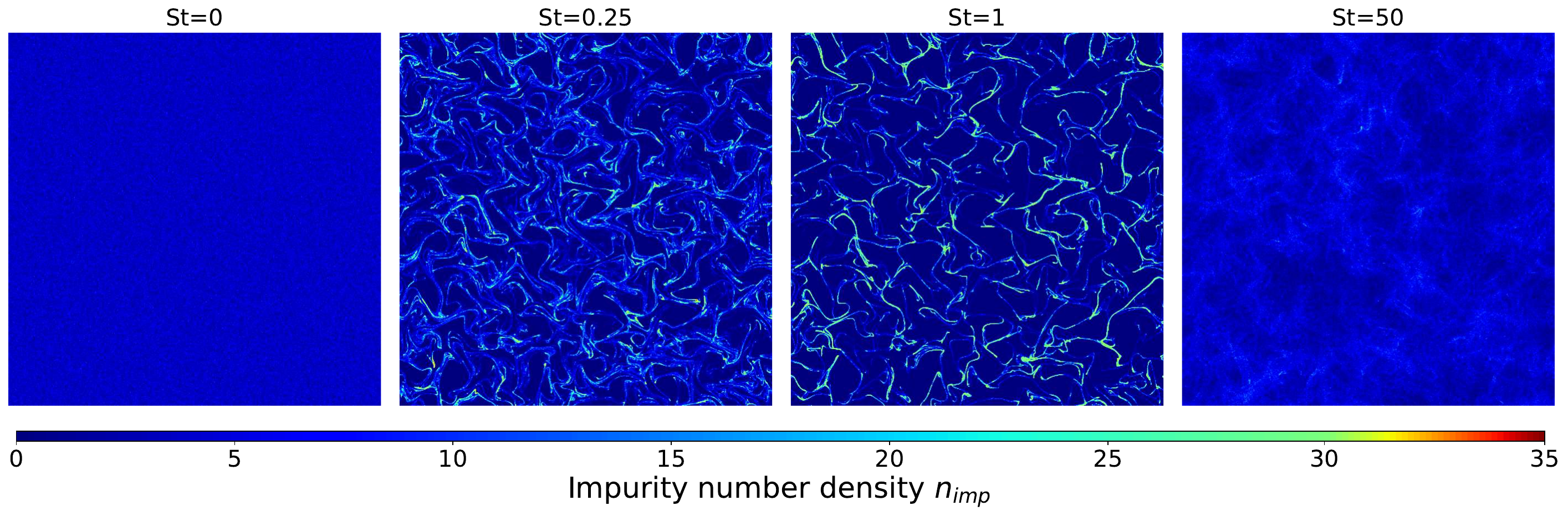}}
        \caption[]{Visualization of impurity density fields ($10^6$ impurity particles) for various Stokes numbers in the classical Hasegawa--Wakatani model ($c = 0.7$). }
        \label{fig: impurity density fields: cHW, c =0.7}
 \end{figure*}

\section{\label{sec: Neural Networks}Neural Networks for Synthesizing Preferential Concentration of Particles}
 
Simulating flow without impurity particles to obtain the vorticity field is not computationally expensive, but including and tracking $10^6$ impurity particles increases costs.  Since a specific statistically stationary distribution of impurity density is linked to a specific vorticity field, we propose creating a neural network that can directly predict the distribution of impurity density based on the vorticity field.  This network uses as input the vorticity field (easily obtained from DNS) and produces as output the corresponding impurity density field for a specific Stokes number. With this neural network, we can run DNS to get the vorticity field,  use it as input for  the network, and efficiently predict the impurity density distribution for a specific Stokes number. This approach eliminates the need for costly simulations involving millions of particles.

 We will evaluate and compare three different neural network architectures that have previously been utilized in particle-laden hydrodynamic turbulence studies:\cite{Oujia2022, Maurel2023NN} the Autoencoder, U-Net, and Generative Adversarial Network (GAN). To determine the most effective neural network architecture for accurately predicting the impurity density fields, a thorough analysis of the statistical attributes of the impurity density fields generated by these neural networks is carried out.
 In the following, we will consider the cHW model for the quasi-adiabatic case ($c = 0.7$).

\subsection{Different machine learning models: Autoencoder, U-Net, and GAN}

\begin{figure*}[htb]
        \centerline{\includegraphics[width=1.0\linewidth]{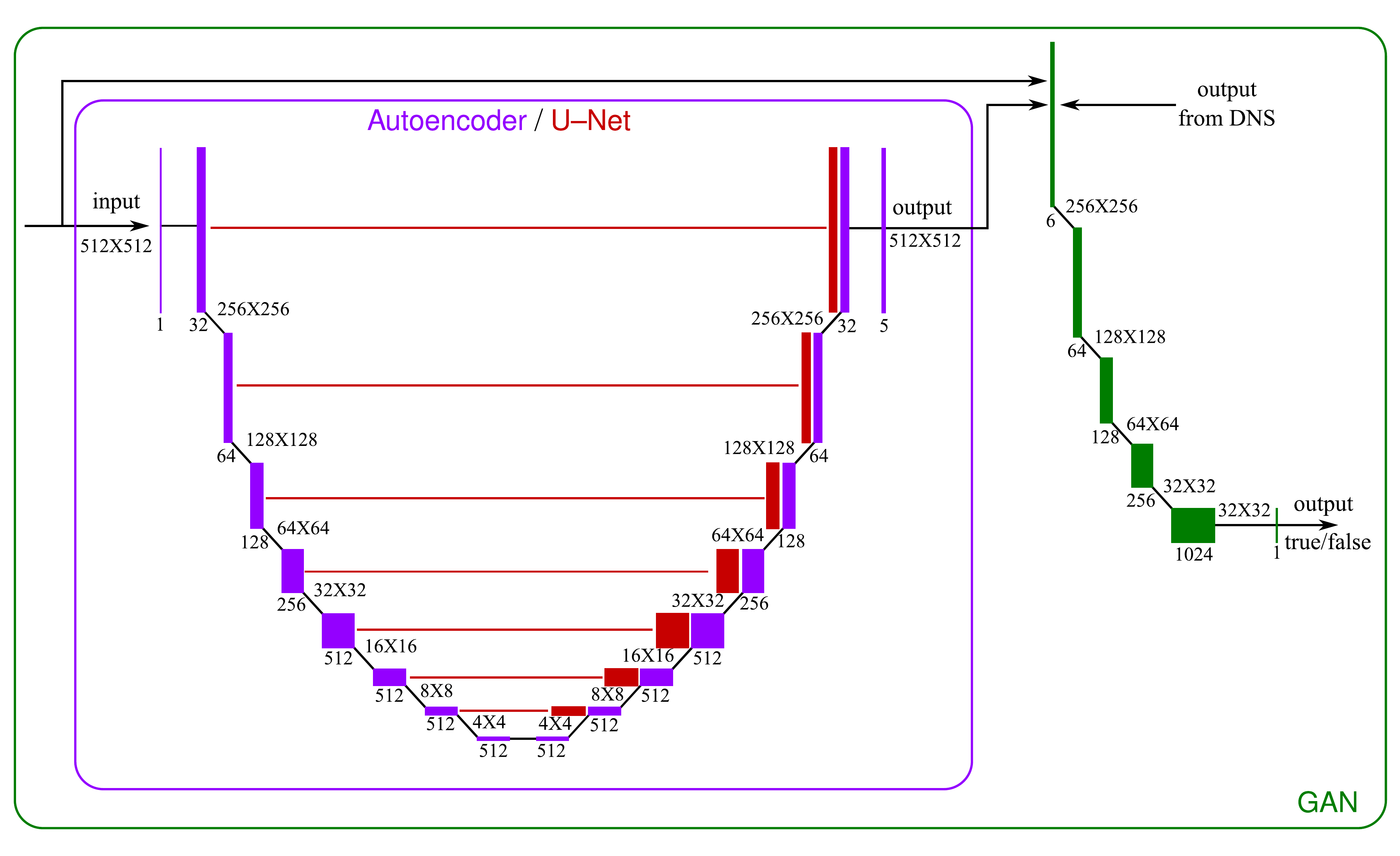}}
        \caption[]{The neural network architectures are illustrated as: Autoencoder in purple; U-Net in purple and red; GAN in purple, red, and green.}
        \label{fig: network architectures}
\end{figure*}

The models used in this study are based on those utilized in Ref.~\onlinecite{Oujia2022} and \onlinecite{Maurel2023NN}. Fig.~\ref{fig: network architectures} displays the architectures tailored from the standard structure, which we have implemented using TensorFlow.
\begin{itemize}
    
    \item \textbf{Autoencoder:} Autoencoder \cite{Rumelhart1986} is the basic model in our study, comprising of an encoder that compresses the data, and a decoder that reconstructs the output from this compressed version. 
    \item \textbf{{U-Net:}} U-Net \cite{Ronneberger2015} is essentially an Autoencoder but with added skip connections. These connections facilitate non-sequential connections between layers which helps in the preservation of information at different scales throughout the network. 
    \item \textbf{GAN:} GAN (Generative Adversarial Network) consists of two neural networks,\cite{Goodfellow2020} a Generator and a Discriminator, trained simultaneously through adversarial processes. The Generator attempts to produce synthetic data, while the Discriminator tries to distinguish between real and synthetic data.  We use U-Net as the generator.\cite{Isola2017} More details can be found in  Ref.~\onlinecite{Oujia2022}  and \onlinecite{Maurel2023NN}.

\end{itemize}

In this study, we use the data of one long simulation in the statistically stationary regime. This simulation has a temporal span equivalent to 220 eddy turnover times. Within this simulation, we have captured 400 snapshots of the vorticity field.  For each snapshot of the vorticity field, there are 5 corresponding impurity density fields ($St = 0.01, 0.05, 0.25, 1, 5$). The neural network is then trained to learn the relation between the vorticity field and the corresponding impurity density field.  70\% of the dataset is for training the model and 30\% for evaluating its performance. Vorticity data are normalized between -1 and 1. Impurity density data are normalized between 0 and 1 to enable the use of binary cross-entropy.

The Adam optimizer \cite{KingmaBa2014}  was used to train the neural networks. Training on a single sample from the training set is termed a `step'. The Autoencoder and U-net models were trained for 28,000 training steps. The GAN model required more steps for convergence and was trained for  56,000 steps The Python codes were run on an NVIDIA graphics card Tesla V100 32GB.

\subsection{Visual Results}

\begin{figure*}[htb]
    \centerline{\includegraphics[width=1.\linewidth]{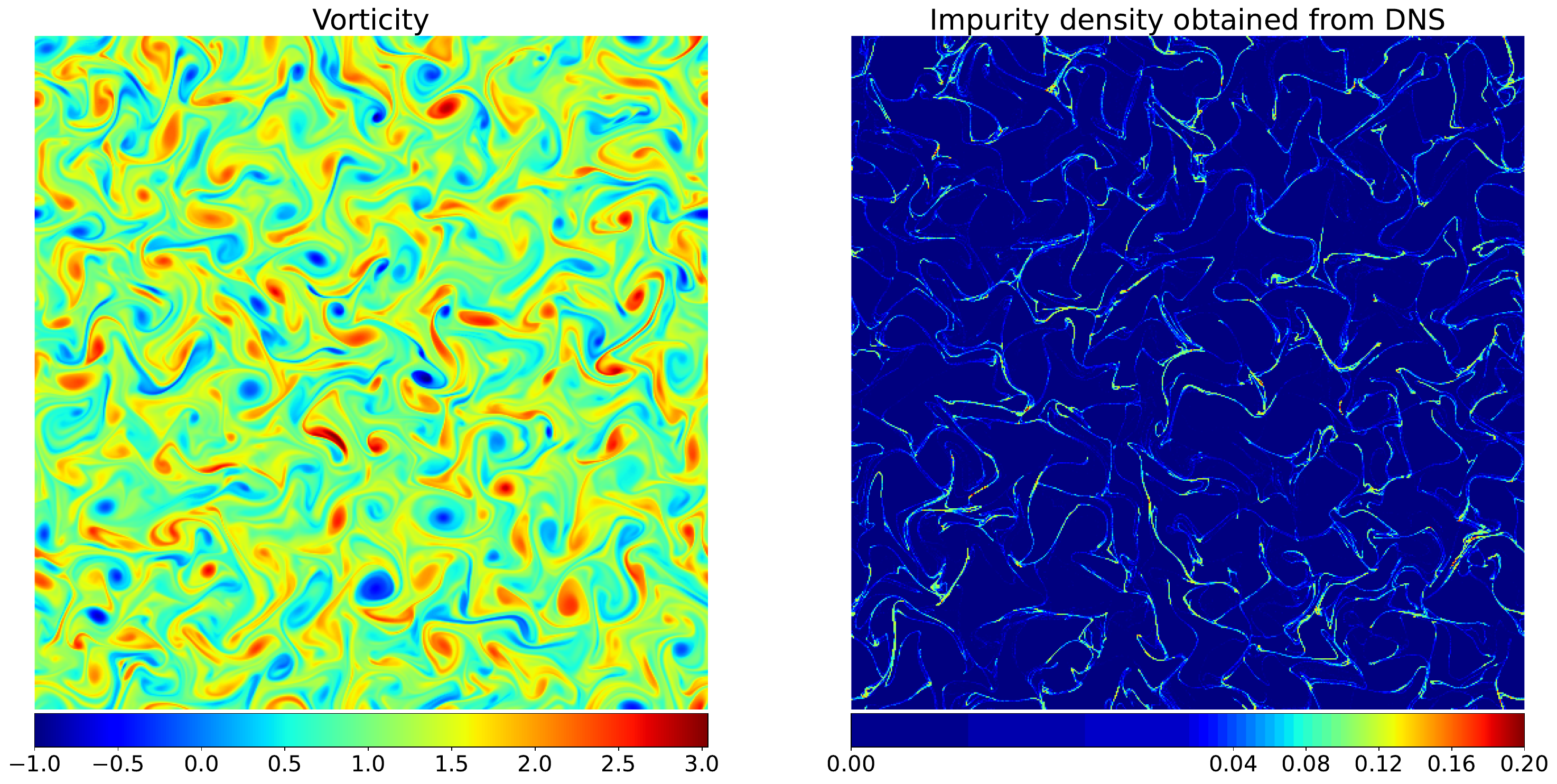}}
        \caption[]{Vorticity field (left)) and corresponding impurity density field (right) ($St  = 1$) obtained from DNS  for the case of $c = 0.7$, cHW.}
        \label{fig: C0_7_input_and_ground_truth_St1.0000}
 \end{figure*}

\begin{figure*}[htb]
    \centerline{\includegraphics[width=1.0\linewidth]{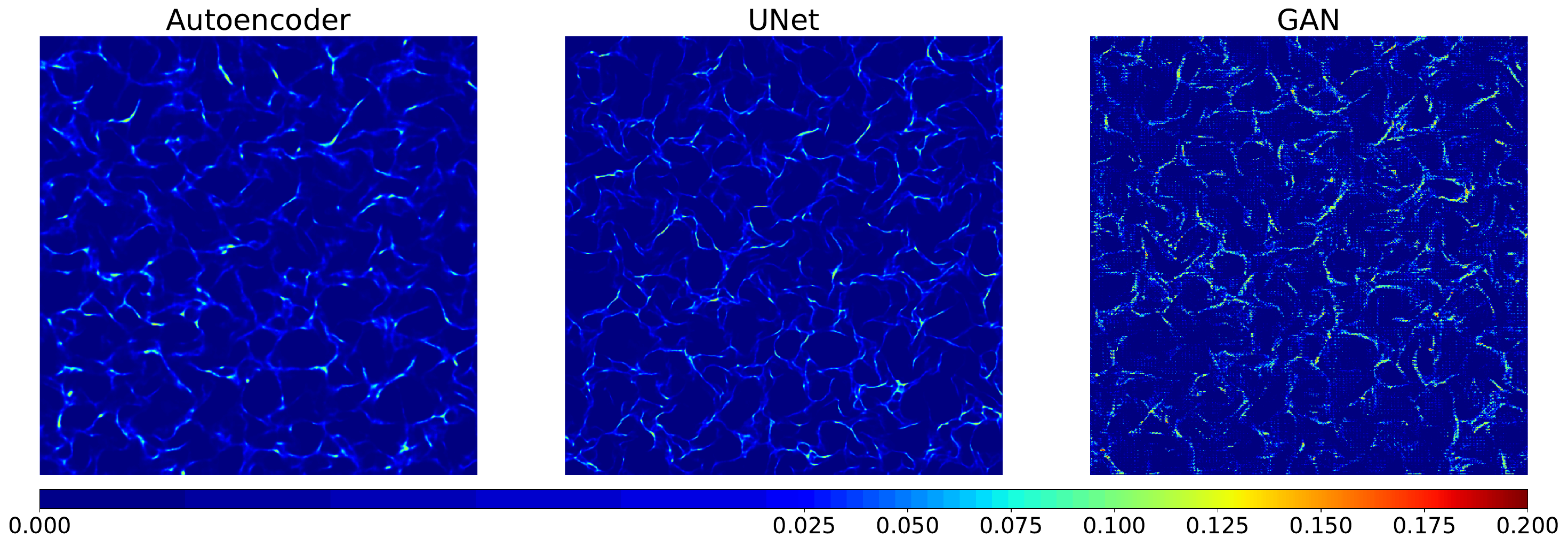}}
        \caption[]{From left to right: For the case of $c = 0.7$ (cHW), impurity density field ($St = 1$)  predicted from Autoencoder, U-Net and GAN, respectively.}
        \label{fig: C0_7_decoded_images_St1.0000}
 \end{figure*}

Fig.~\ref{fig: C0_7_input_and_ground_truth_St1.0000} shows the vorticity field and corresponding impurity density field ($St  = 1$) obtained from DNS for the case of $c = 0.7$ (cHW). Fig.~\ref{fig: C0_7_decoded_images_St1.0000} illustrates impurity density field ($St = 1$)  predicted from Autoencoder, Unet and GAN for $c = 0.7$ (cHW) using vorticity field as input. When comparing DNS impurity density fields with those predicted by the Autoencoder and U-Net, we find that both DNS and predicted fields exhibit similar structures, but the clustering in the predicted fields is slightly fuzzy.  The GAN model successfully predicts void areas and filament-like structures to some degree of success, but it sometimes struggles with predicting the details of regions with impurities.

\subsection{Statistical analysis}

\begin{figure}[htb]
      \centerline{\includegraphics[width=0.5\textwidth]{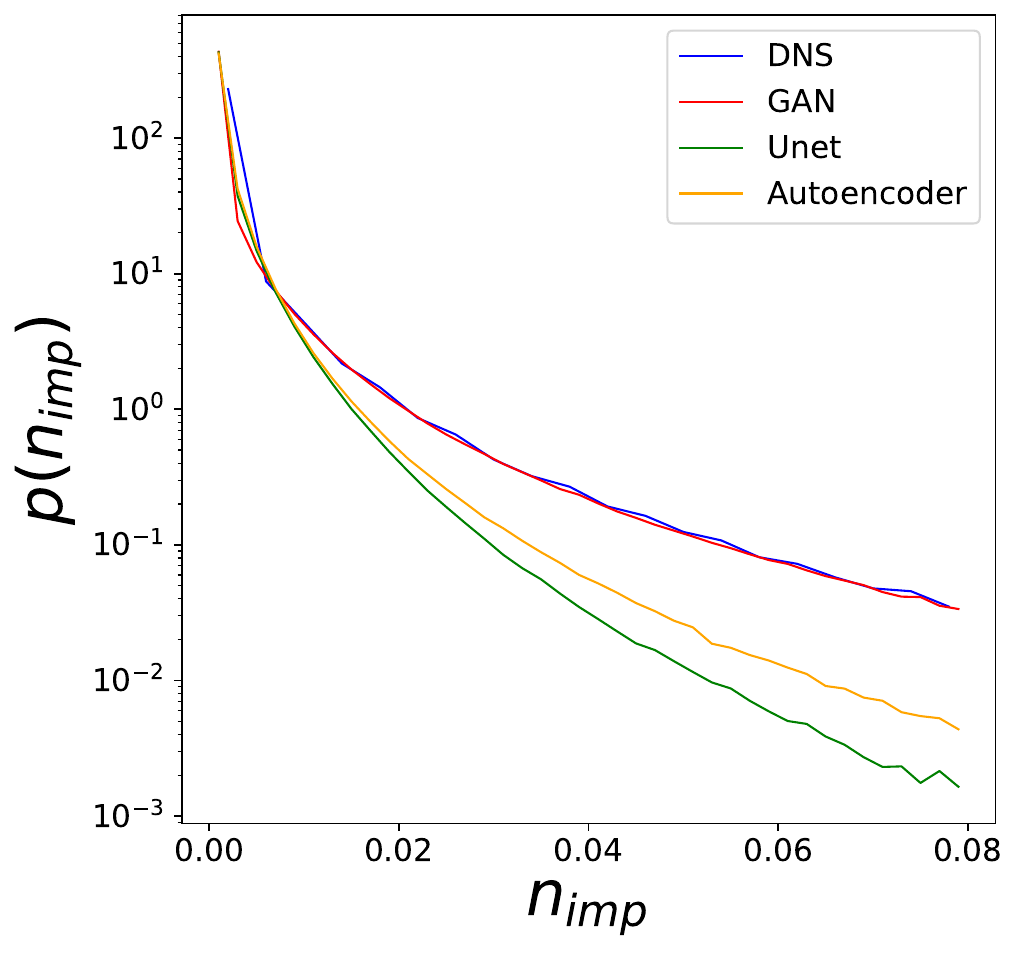}}
        \caption[]{PDFs of impurity density of DNS data and predicted data using three different architectures for $St = 1$.}
        \label{fig: PDFs of impurity density for St = 1}
 \end{figure}

To better understand the performance of our models, we analyze the results using the probability density functions (PDFs) and energy spectra of the impurity number density. Fig.~\ref{fig: PDFs of impurity density for St = 1} shows the PDFs of DNS data and predicted data using three different architectures for $St = 1$.  From Fig.~\ref{fig: PDFs of impurity density for St = 1} we observe that the GAN model excels in predicting the density distribution, as its predictions closely align with DNS. In contrast, U-Net and Autoencoder exhibit shortcomings, particularly in predicting higher densities.  The deviations in the PDFs of U-Net and the Autoencoder suggest that these models may struggle to accurately replicate the intricate characteristics present in high-density regions.

\begin{figure}[htb]
      \centerline{\includegraphics[width=0.5\textwidth]{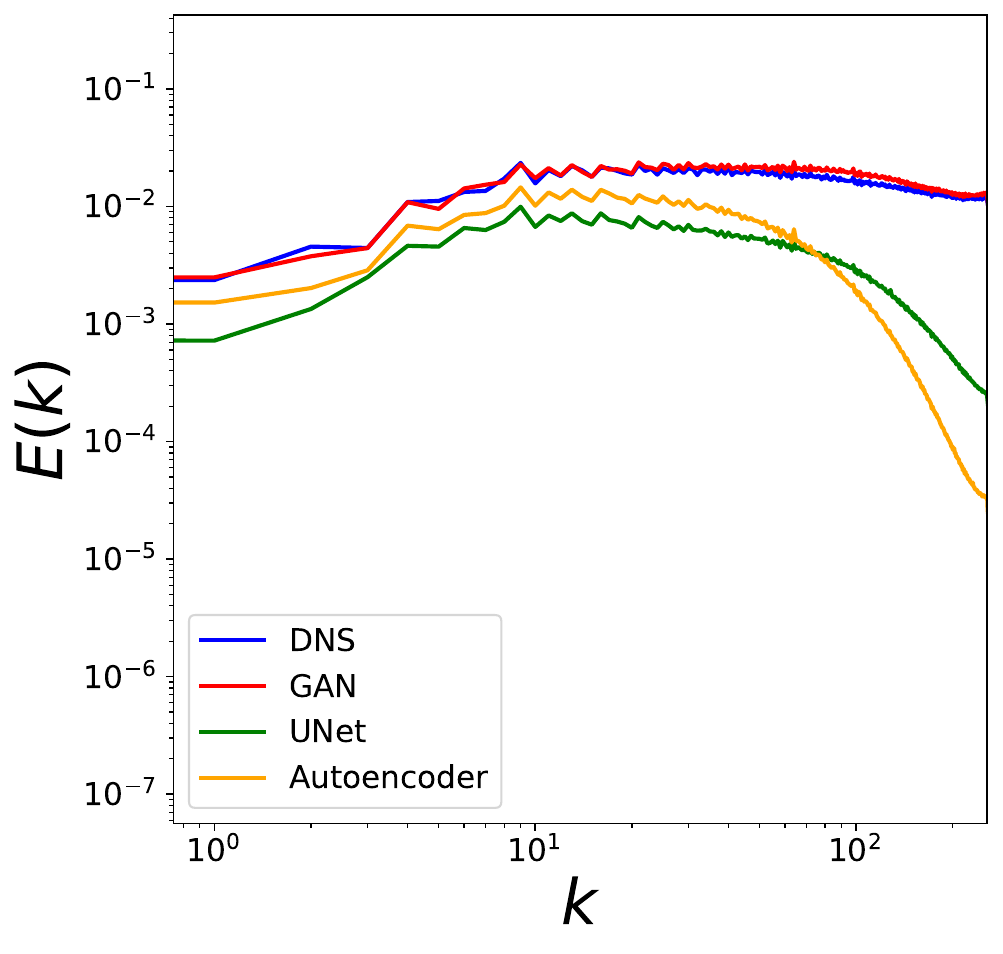}}
        \caption[]{Energy spectra of impurity density of DNS data and predicted data using three different architectures for $St = 1$.}
        \label{fig: energy of impurity density for St = 1}
 \end{figure}

Fig.~\ref{fig: energy of impurity density for St = 1} shows the energy spectra of DNS data and predicted data using three different architectures for $St = 1$. In Fig.~\ref{fig: energy of impurity density for St = 1}  we observe that the GAN's energy spectra are nearly identical to that of the DNS data. This demonstrates the remarkable accuracy of the GAN model in predicting the energy spectra of the impurity density. On the contrary, both U-Net and the Autoencoder exhibit a significant drop at high wave numbers, which correspond to fine structures in the data. This indicates that U-Net and the Autoencoder are unable to accurately capture these fine-scale structures and may struggle to represent the intricate details present in the impurity density.

\begin{figure}[htb]
      \centerline{\includegraphics[width=0.5\textwidth]{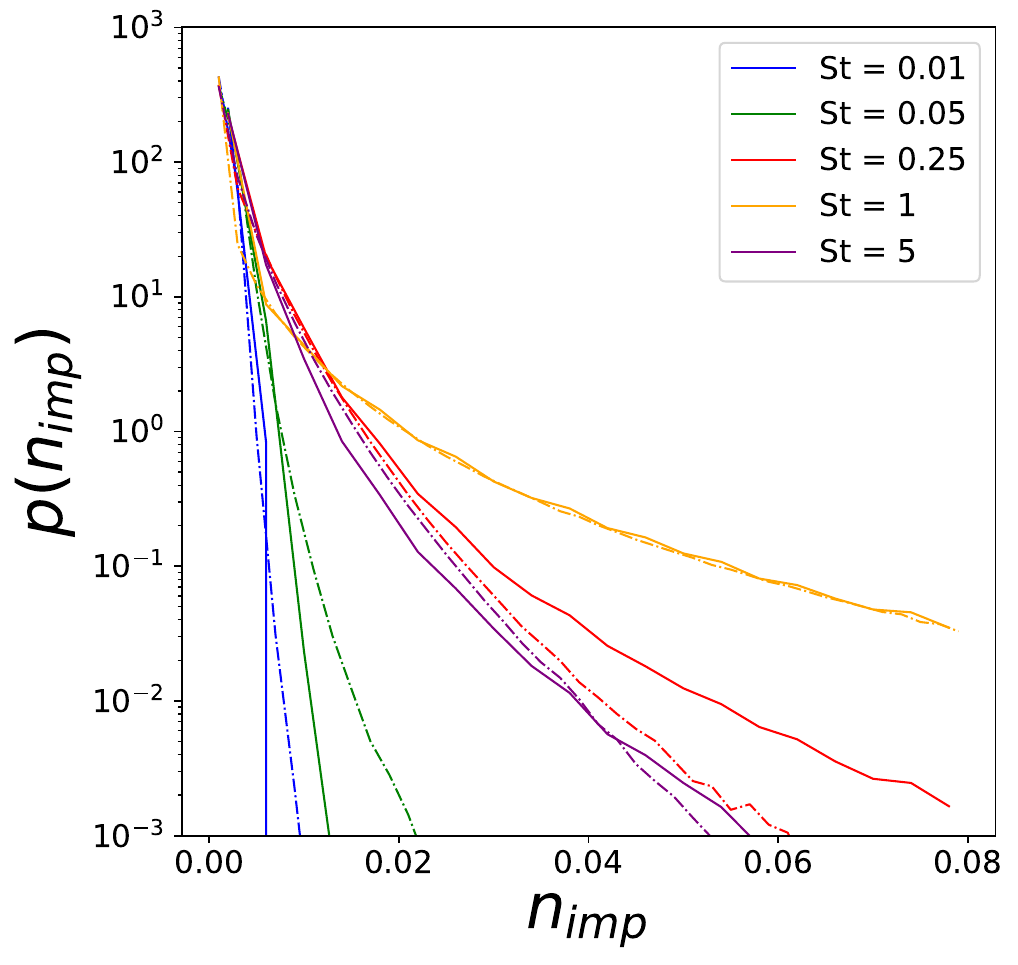}}
        \caption[]{PDFs of impurity density of DNS data (solid lines) and predicted data using GAN (dashed lines) for different $St$.}
        \label{fig: PDFs of impurity density for GAN}
 \end{figure}

\begin{figure}[htb]
        \centerline{\includegraphics[width=0.5\textwidth]
        {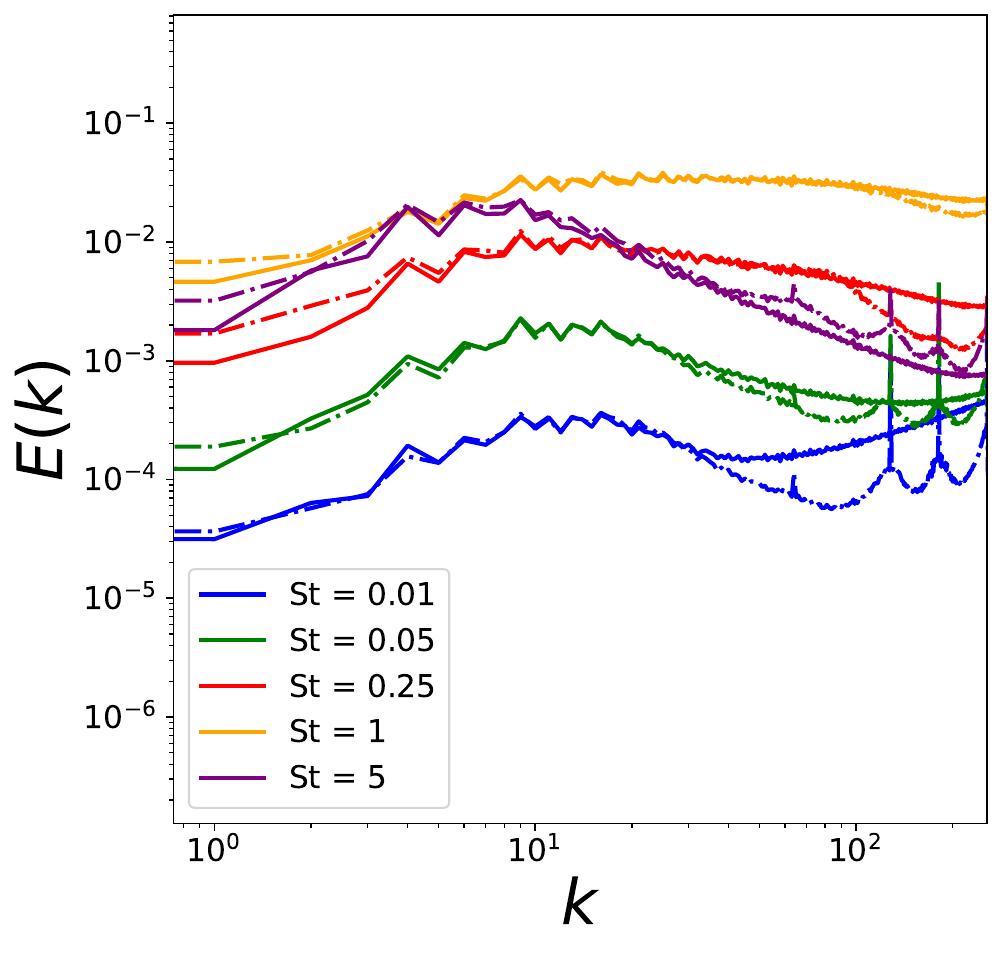}}
        \caption{Energy spectra of impurity density of DNS data (solid lines) and predicted data using GAN (dashed lines) for different $St$.}
        \label{fig: energy of impurity density for GAN}
\end{figure}

Overall, the analysis of the PDFs and energy spectra reaffirms that GAN outperforms the other two for $St = 1$. Thus, we utilize GAN in the following to investigate more Stokes numbers values. From the PDF of impurity density as shown in Fig.~\ref{fig: PDFs of impurity density for GAN}, we can see that for $ St = 0.25$, GAN predicts lower impurity density when compared with DNS. For small $St$ ($St = 0.01, 0.05$), GAN predicts more particles than DNS. When dealing with a large Stokes number ($St = 5$), the PDF is similar to the DNS results. In the case of $St =1$, the GAN predictions align perfectly with DNS. For energy spectra, as shown in Fig.~\ref{fig: energy of impurity density for GAN}, it is noticeable that the GAN predictions align very well with the DNS. In the GAN predicted energy spectra, notable spikes were identified at frequencies corresponding to the edges and corners of the domain. This could be because the initial encoding of domain periodicity was inadequate. To correct this, adjustments in the code are needed for true periodic convolution. Unfortunately, as of the time of this writing, TensorFlow does not yet provide native support for periodic padding.

In our study, we train the AI model specifically for $c=0.7$, which is a typical value for edge plasma.  This model predicts impurity density fields for $St = 0.01, 0.05, 0.25, 1$ and $5$  using the vorticity field at $c = 0.7$ as input.  The model's validity is specific to $c = 0.7$  and the aforementioned Stokes numbers. Its extrapolability is shown and discussed in Appendix~\ref{appendix: extrapolability}.

Now we know that GAN is the optimal model. Next, we can conduct simulations to generate more vorticity fields without particles, a process which is computationally efficient. We then input these vorticity fields into GAN to predict the impurity density fields. Subsequently, we can perform detailed analyses of transport characteristics and clustering, using methods like multiresolution analysis and Voronoi tessellation.  There are several possible ways to enhance the model. One method is to adjust architectural features like the number of layers, filter sizes, and channel counts. Another effective approach is to integrate physical constraints into the model, ensuring that predictions align with established physical laws, such as mass or energy conservation, for more accurate and plausible results.

\section{\label{sec: Conclusions}Conclusions}
This study focuses on inertial impurity within fusion plasmas, which is different from previous literature where the focus has typically been on `passive tracers' following the flow. We used high-resolution numerical simulations based on Hasegawa--Wakatani equations with impurity particles having different inertia.  We studied the quasi-adiabatic regime ($c = 0.7$),  typical for Tokamak edge plasma.  Our findings show that the inertia of particles, quantified by the Stokes number $St$, plays a vital role in their behavior. At lower $St$ values, impurities move along the flow streamlines. As $St$ increases, impurities tend to cluster at low-vorticity regions. At very high $St$ values, impurities show random movement due to substantial inertia, resulting in much less clustering.

As simulating $10^6$ or more impurity particles is a computationally intensive task, we utilized machine learning techniques to create a surrogate model.  Three neural networks - Autoencoder, U-Net, and GAN - were developed to synthesize impurity densities, using vorticity fields as input. All models produced visually comparable results to simulations, with varying degrees of success. The GAN closely matches DNS values in terms of probability density functions and energy spectra of density. This confirms GAN's effectiveness in modeling both the general distribution and finer structures. Based on the results, future work will involve: Improving GAN by adjusting the architectures or involving physical constraints directly into the model; running simulations for more vorticity fields, then using the GAN to predict impurity fields for analyzing clustering and transport properties; Including electromagnetic interactions for more accurate impurity particle models.

\begin{acknowledgments}
ZL, TMO and KS acknowledge the financial support from I2M and the French Federation for Magnetic Fusion Studies (FR-FCM) and the Eurofusion consortium,  funded by the  Euratom  Research and Training Programme under Grant Agreement No. 633053. The views and opinions expressed herein do not necessarily reflect those of the European Commission. ZL, TMO, BK, PK and KS acknowledge partial funding from the Agence Nationale de la Recherche (ANR), project CM2E, grant ANR-20-CE46-0010-01.
Centre de Calcul Intensif d’Aix-Marseille is acknowledged for providing access to its high performance computing resources.
\end{acknowledgments}

\section*{Data Availability Statement}
The data that support the findings of this study are available from the corresponding author upon reasonable request.
%----------------------------------------------------------------------------
\appendix

\section{\label{appendix: zonal flows}Impurities in zonal flows}

\begin{figure}[htb]
        \centerline{\includegraphics[width=1.0\linewidth]{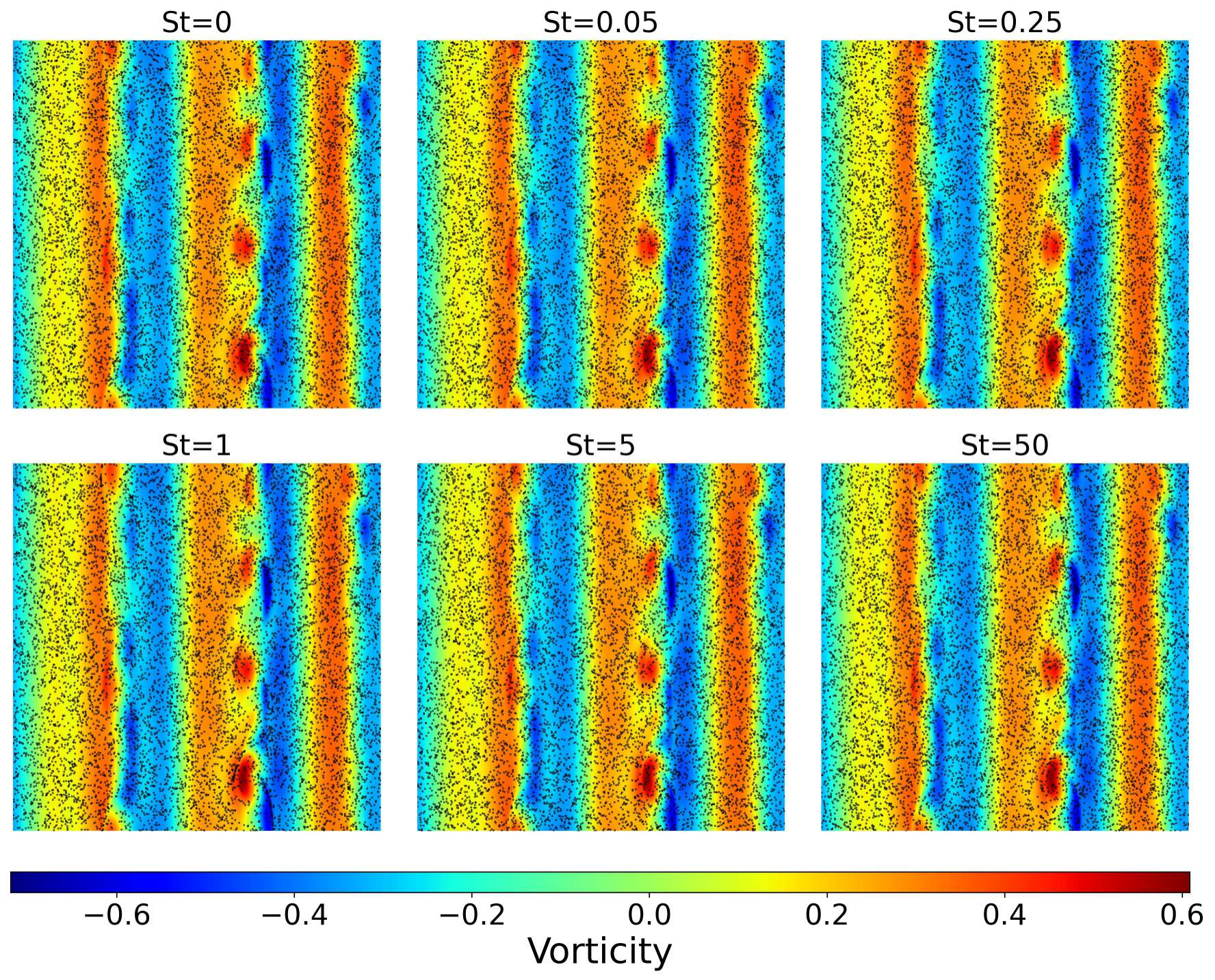}}
        \caption[]{Visualization of vorticity fields in fully developed turbulence regime and $10^4$ superimposed impurity particles for various Stokes numbers in the modified Hasegawa--Wakatani model ($c = 2$).}
        \label{fig: vorticity and particles: mHW, c =2}
 \end{figure}

\begin{figure}[htb]
        \centerline{\includegraphics[width=1.0\linewidth]{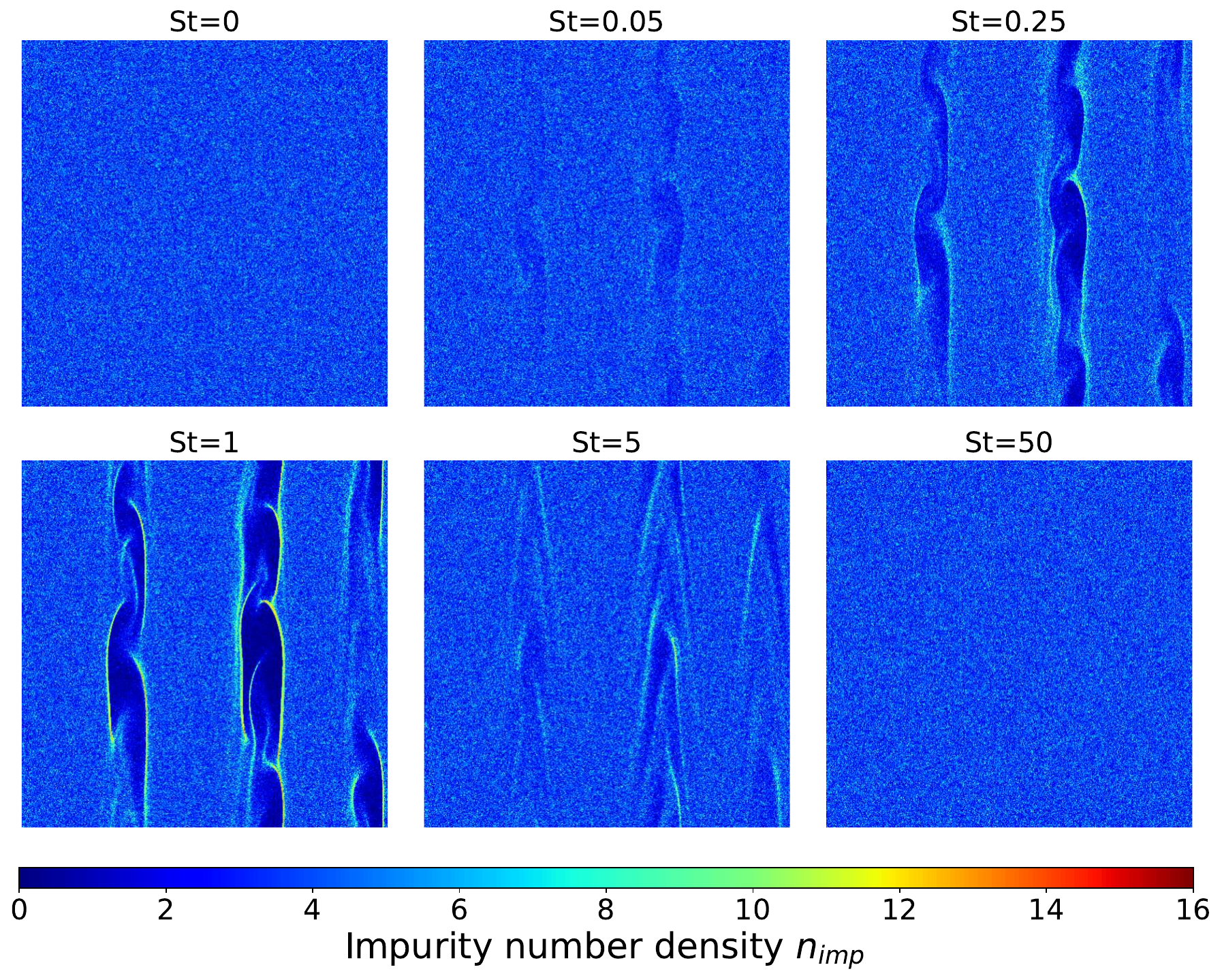}}
        \caption[]{Visualization of impurity density fields ($10^6$ impurity particles) for various Stokes numbers in the modified Hasegawa--Wakatani model ($c = 2$).}
        \label{fig: impurity density fields: mHW, c =2}
 \end{figure}
In the modified Hasegawa--Wakatani model at $c=2$ with $10^4$ impurity particles as shown in Fig. \ref{fig: vorticity and particles: mHW, c =2}, the zonal flow is very strong and the vorticity values are weaker compared to cHW. The particles appear to be distributed evenly across the fluid, and in proximity to the zonal flow, it seems that clustering is not evident To gain deeper insights, we examined the impurity density fields obtained from $10^6$ impurity particles. As depicted in Fig.~\ref{fig: impurity density fields: mHW, c =2}, there is some preferential concentration of impurities in the shear layer region for $St=0.25$, 1 and 5. The impurity particles are ejected by the zonal flow, and for St=1 the effect seems to be the strongest. For $St =0$ and 50 the impurity particles are randomly distributed.

\section{\label{appendix: extrapolability}Extrapolability}
In the following we test the extrapolability of the GAN model for different $c$-values. The model has been trained using the data with $c = 0.7$. As input data we then use vorticity fields for $c$-values different from 0.7 and assess the impurity densities obtained with the GAN model. The results are shown from Fig.~\ref{fig: PDFs of impurity density for GAN, c = 0.01} to Fig.~\ref{fig: energy of impurity density for GAN, c = 2}. In the hydrodynamic regime ($c = 0.01$), we observe that GAN predicts the PDF and  energy spectra of impurity density very well at $St  =1$. At other Stokes numbers, discrepancies arise between GAN predictions and DNS results in terms of PDFs and energy spectra. In  the adiabatic regime ($c = 2$), the PDFs and energy spectra of impurity density predicted by GAN tend to have larger values  than the DNS results. Note that the results are good from statistical point of view, but the obtained impurity density fields differ visually from the DNS results.

\begin{figure}[htb]
      \centerline{\includegraphics[width=0.5\textwidth]{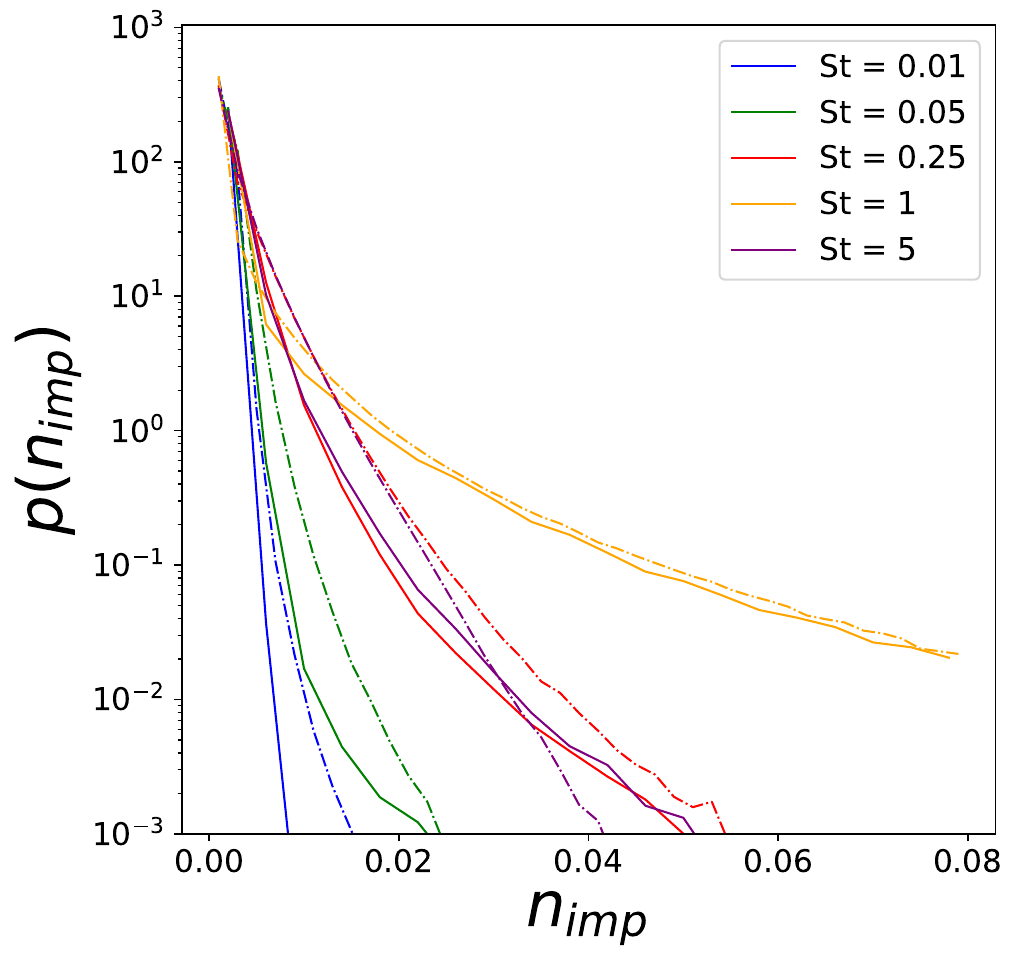}}
        \caption[]{PDFs of impurity density of DNS data (solid lines) and predicted data using GAN (dashed lines) for different $St$ in the hydrodynamic regime ($c = 0.01$).}
        \label{fig: PDFs of impurity density for GAN, c = 0.01}
 \end{figure}

\begin{figure}[htb]
        \centerline{\includegraphics[width=0.5\textwidth]{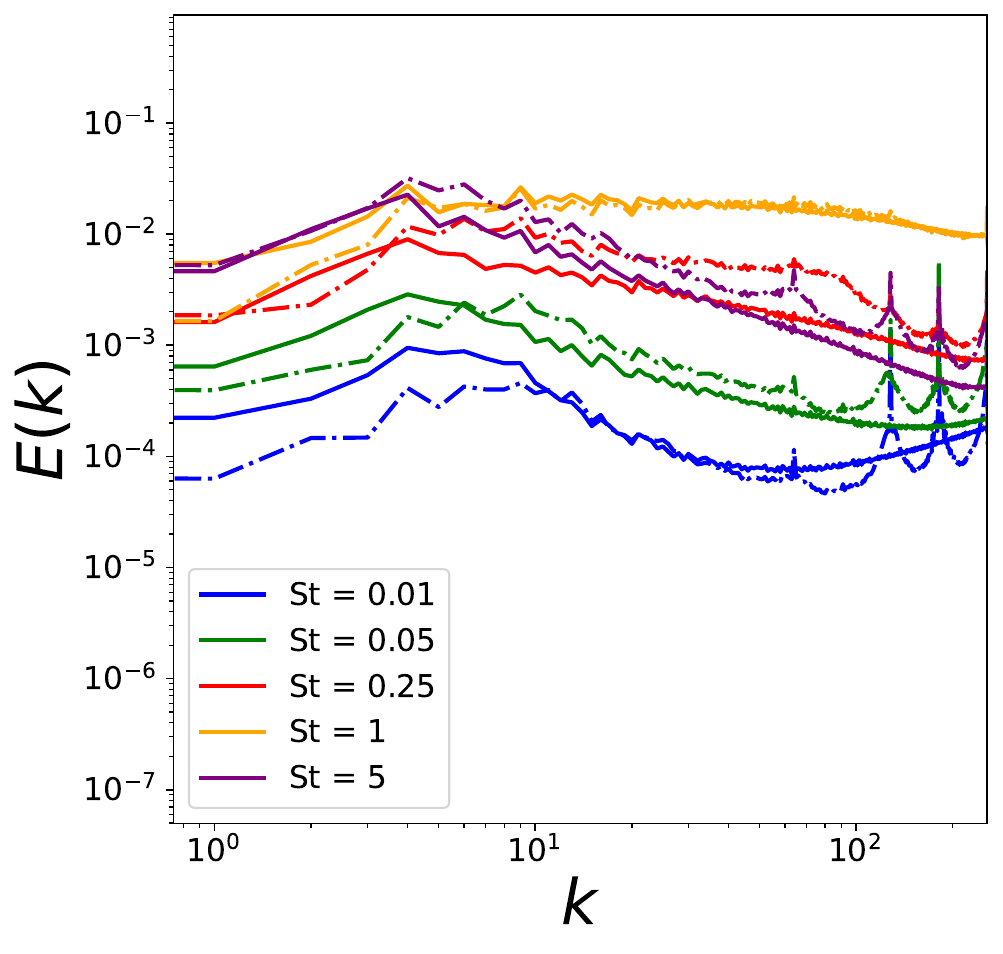}}
        \caption{Energy spectra of impurity density of DNS data (solid lines) and predicted data using GAN (dashed lines) for different $St$ in hydrodynamic regime ($c = 0.01$).}
        \label{fig: energy of impurity density for GAN, c = 0.01}
\end{figure}

\begin{figure}[htb]
      \centerline{\includegraphics[width=0.5\textwidth]{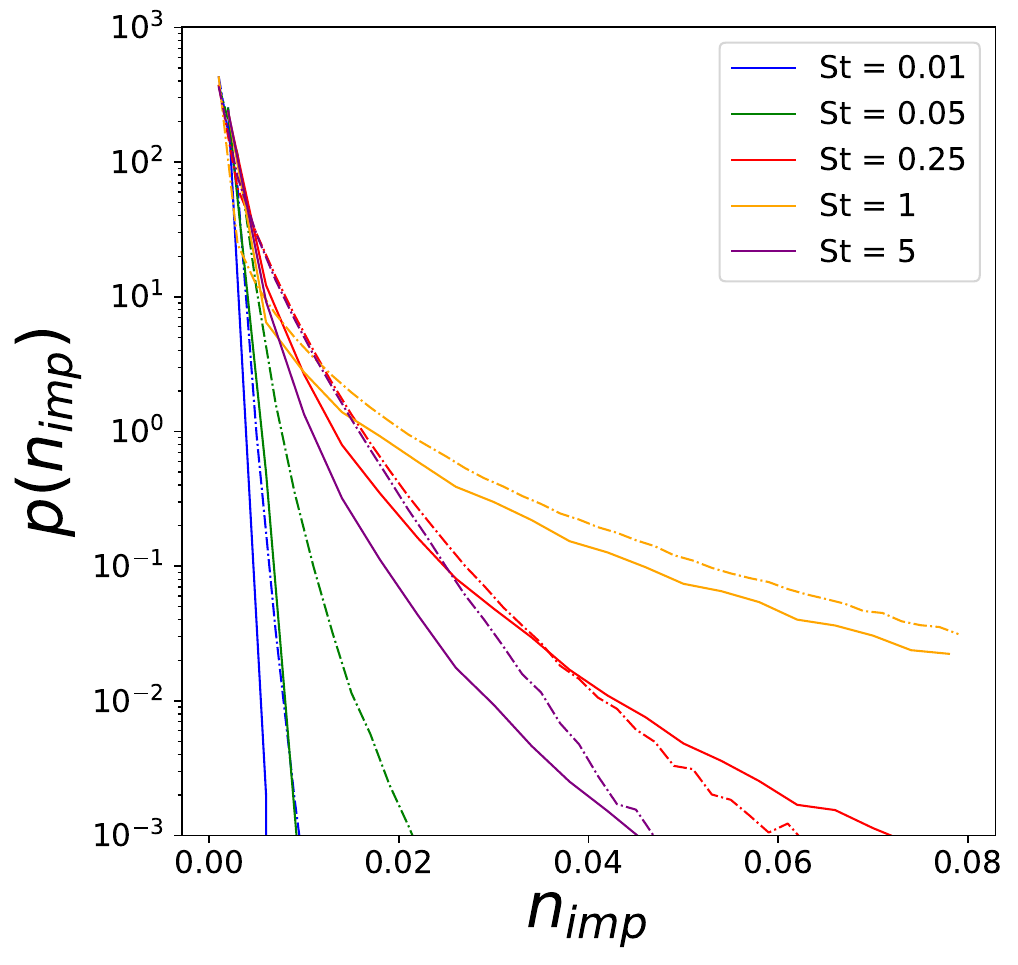}}
        \caption[]{PDFs of impurity density of DNS data (solid lines) and predicted data using GAN (dashed lines) for different $St$ in the adiabatic regime ($c = 2$).}
        \label{fig: PDFs of impurity density for GAN, c = 2}
 \end{figure}

\begin{figure}[h!]
        \centerline{\includegraphics[width=0.5\textwidth]{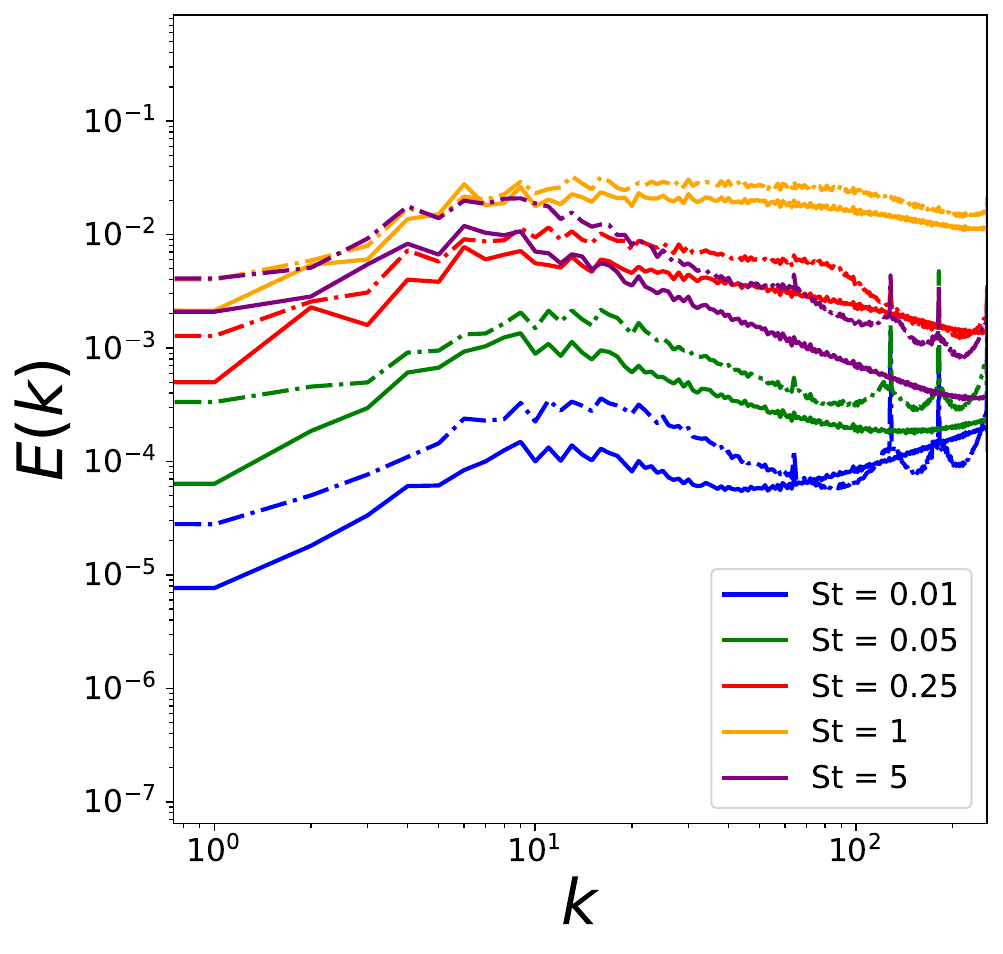}}
        \caption{Energy spectra of impurity density of DNS data (solid lines) and predicted data using GAN (dashed lines) for different $St$ in the adiabatic regime ($c = 2$).}
        \label{fig: energy of impurity density for GAN, c = 2}
\end{figure}

\nocite{*}

%\bibliography{aipsamp}% Produces the bibliography via BibTeX.

\end{document}